\documentclass[preprint]{aastex}
\newcommand{\Fermi}{\emph{Fermi}}
\newcommand{\g}{$\gamma$}

\begin{document}
\title{Discovery of Pulsed \g-rays from PSR J0034$-$0534 with the \Fermi{} LAT:  A Case for Co-located Radio and \g-ray Emission Regions}
\author{
A.~A.~Abdo\altaffilmark{2,3}, 
M.~Ackermann\altaffilmark{4}, 
M.~Ajello\altaffilmark{4}, 
A.~Allafort\altaffilmark{4}, 
L.~Baldini\altaffilmark{5}, 
J.~Ballet\altaffilmark{6}, 
G.~Barbiellini\altaffilmark{7,8}, 
D.~Bastieri\altaffilmark{9,10}, 
K.~Bechtol\altaffilmark{4}, 
R.~Bellazzini\altaffilmark{5}
% ...
}
% ...
\altaffiltext{2}{Space Science Division, Naval Research Laboratory, Washington, DC 20375, USA}
\altaffiltext{3}{National Research Council Research Associate, National Academy of Sciences, Washington, DC 20001, USA}
\altaffiltext{4}{W. W. Hansen Experimental Physics Laboratory, Kavli Institute for Particle Astrophysics and Cosmology, Department of Physics and SLAC National Accelerator Laboratory, Stanford University, Stanford, CA 94305, USA}
\altaffiltext{5}{Istituto Nazionale di Fisica Nucleare, Sezione di Pisa, I-56127 Pisa, Italy}
\altaffiltext{6}{Laboratoire AIM, CEA-IRFU/CNRS/Universit\'e Paris Diderot, Service d'Astrophysique, CEA Saclay, 91191 Gif sur Yvette, France}
\altaffiltext{7}{Istituto Nazionale di Fisica Nucleare, Sezione di Trieste, I-34127 Trieste, Italy}
\altaffiltext{8}{Dipartimento di Fisica, Universit\`a di Trieste, I-34127 Trieste, Italy}
\altaffiltext{9}{Istituto Nazionale di Fisica Nucleare, Sezione di Padova, I-35131 Padova, Italy}
\altaffiltext{10}{Dipartimento di Fisica ``G. Galilei", Universit\`a di Padova, I-35131 Padova, Italy}
% ...

\begin{abstract}
Millisecond pulsars (MSPs) have been firmly established as a class of \g-ray emitters via the detection of pulsations above 0.1 GeV from eight MSPs by the \Fermi{} Large Area Telescope (LAT).  Using thirteen months of LAT data significant \g-ray pulsations at the radio period have been detected from the MSP PSR J0034$-$0534, making it the ninth clear MSP detection by the LAT.  The \g-ray light curve shows two peaks separated by 0.274$\pm$0.015 in phase which are very nearly aligned with the radio peaks, a phenomenon seen only in the Crab pulsar until now.  The $\geq$0.1 GeV spectrum of this pulsar is well fit by an exponentially cutoff power law with a cutoff energy of 1.8$\pm$0.6$\pm$0.1 GeV and a photon index of 1.5$\pm$0.2$\pm$0.1, first errors are statistical and second are systematic.  The near-alignment of the radio and \g-ray peaks strongly suggests that the radio and \g-ray emission regions are co-located and both are the result of caustic formation.
\end{abstract}

\keywords{{\it Facilities:} \facility{Fermi ()}\ \ {\it Pulsars:} gamma-ray, individual (PSR J0034$-$0534)}

\section{INTRODUCTION}
Forty new pulsars have recently been observed to pulse in high-energy (HE), $\geq$0.1 GeV, \g-rays by the Large Area Telescope (LAT) aboard the \emph{Fermi Gamma-ray Space Telescope} (formerly GLAST) \citep{psrcat}.  Among these new detections are eight millisecond pulsars (MSPs) \citep{abdoa}.  MSPs are thought to be older, recycled pulsars in binary systems \citep{alpar}, a theory which has been supported by the discovery of millisecond pulsations from accreting low-mass X-ray binaries (LMXBs) (e.g. Wijnands \& van der Klis 1998) and more recently by observations of an MSP transitioning from the LMXB to the pulsar phase \citep{arch}.  \citet{usov} first attempted to show that MSPs should be \g-ray emitters and could, possibly, have an even greater HE luminosity than the pulsar in the Crab nebula.  Data from the \emph{Energetic Gamma-Ray Experiment Telescope} (EGRET) \citep{DJT} were searched for both point source and pulsed emission from several MSPs \citep{fierro} and, while none were detected, upper limits were set which put useful constraints on MSP emission models.  \citet{0218} did report a marginal ($\sim4\sigma$) pulsed detection of the MSP PSR J0218+4232, which has been confirmed with \Fermi{} \citep{abdoa}.  HE \g-ray emission has also been detected from the vicinity of the globular cluster 47 Tucanae with a \g-ray spectrum consistent, at the 95\% confidence level, with the superposition of between seven and sixty-two MSPs \citep{abdob}.  The Italian observatory \emph{AGILE} has also reported a 4.2$\sigma$ detection of \g-ray pulsations from PSR B1821$-$24 in the globular cluster M28 \citep{AGILEpsrs}.

\Fermi{} began nominal sky-survey observations on 2008 August 4, viewing the entire sky every two orbits ($\sim$3 hours).  The main instrument on \Fermi{} is the Large Area Telescope (LAT), which is a pair-conversion telescope sensitive to \g-rays with energies from 0.02 to $>$300 GeV \citep{LAT}.  The LAT has a 2.4 sr field of view, a peak effective area of $\sim8000\ \rm{cm}^{2}$ above 1 GeV on axis, and a 68\% containment radius of $0.6^{\circ}$ at 1 GeV for events converting in the front section of the LAT.  The LAT timing is derived from a GPS clock on the spacecraft, and events are time-stamped to an accuracy better than 1 $\mu$s \citep{abdoc}.  Using approximately thirteen months of LAT data, we have discovered \g-ray emission from PSR J0034$-$0534, which thus becomes the ninth \g-ray MSP detected with by \Fermi{} LAT.

\section{PSR J0034$-$0534}
PSR J0034$-$0534 was discovered by \citet{disc} in a survey of the southern sky with the Parkes radio telescope.  The pulsar is in a binary sytem with a 1.6 d orbital period.  The initial observations measured a spin period (P) of 1.877 ms, which makes this the fastest \g-ray MSP yet detected with the LAT, and a period derivative ($\dot{\rm{P}}$) of $6.7\times10^{-21}$.  Table 1 lists some of the measured and derived timing parameters for PSR J0034$-$0534 using the radio data described in \S3.1. The derived quantities in Table 1 are corrected for the Shklovskii effect \citep{shklov} and assume a moment of inertia $\rm{I} = 10^{45}\ \rm{g}\ \rm{cm}^{2}$.  \citet{abdoa} observed that \g-ray MSPs had similar spectral properties and, due to their short periods, comparable magnetic field strengths at the light cylinder ($\rm{B}_{LC}$) to those of younger \g-ray pulsars.  Additionally, \citet{psrcat} noted a weak correlation of cutoff energy with $\rm{B}_{LC}$ for all forty-six \g-ray pulsars detected by the LAT in the first six months.  Of the \g-ray MSPs detected so far only PSR J0218+4232 has a higher value of $\rm{B}_{LC}$ than PSR J0034$-$0534 while most of the younger \g-ray pulsars detected with \Fermi{} have lower values of $\rm{B}_{LC}$.

Large values of $\rm{B}_{LC}$ have been linked to giant pulses (GP) in the radio (Cognard et al. 1996 and Knight et al. 2005) and the relatively large value for PSR J0034$-$0534 prompted several GP searches (Romani \& Johnston 2001, McLaughlin \& Cordes 2003, and Knight et al. 2005) though none were detected. \emph{Hubble Space Telescope} observations revealed an optical counterpart to the binary companion of PSR J0034$-$0534 consistent with a white dwarf hypothesis \citep{opt2}.  Infrared observations of PSR J0034$-$0534 only put upper limits on the flux density of any surronding debris disk (Greaves \& Holland 2000 and Lazio \& Fischer 2004).  \emph{XMM-Newton} observations revealed a low significance ($<3\sigma$) X-ray source 0.2\arcsec{} from the radio position of PSR J0034$-$0534, but no pulsed signal was detected from the source and a firm identification could not be made \citep{xmm}.  The EGRET $3\sigma$ point source flux upper limit, 0.1 to 10 GeV, for PSR J0034$-$0534 of $15.2\times10^{-8}\ \rm{cm}^{-2}\ \rm{s}^{-1}$ \citep{fierro} is an order of magnitude above the LAT flux in \S4.2.

\section{OBSERVATIONS}

\subsection{Radio Timing}
The timing solution used for PSR J0034$-$0534 has been derived from observations carried out at the Nan\c{c}ay radio telescope in France \citep{nancay} and the Westerbork Synthesis Radio Telescope (WSRT) in the Netherlands \citep{wbork2}, using 170 Times Of Arrival (TOAs) recorded between 2005 October 26 and 2009 August 24. Among the 170, 139 were recorded at the WSRT at frequencies between 314 MHz and 376 MHz with a mean uncertainty of 2.5 $\mu$s and a bandwidth of 10 MHz \citep{stappers}. The remaining TOAs were recorded at the Nan\c{c}ay radio telescope at 1398 MHz with a mean uncertainty of 3.2 $\mu$s with a bandwidth of 64 MHz before 2008 June 13 and 128 MHz thereafter \citep{ctdf}. This multiple-frequency dataset tightly constrains the dispersion measure (DM), which is crucial for profile comparisons at different wavelengths. The ephemeris was derived using the TEMPO2\footnote{http://tempo2.sourceforge.net/} pulsar timing package \citep{t2}, fitting for the rotation frequency and its first derivative while accounting for the binary motion of the millisecond pulsar. The combined timing solution gives a post-fit rms of 7.1 $\mu$s and a DM measurement of $13.76517\pm0.00004\ \rm{cm}^{-3}\ \rm{pc}$, with no indication of variation with time. The uncertainty in the dispersion measure leads to an uncertainty of less than 2 $\mu$s in the extrapolation of 300 MHz arrival times to infinite frequency, negligible for the low-statistics gamma-ray lightcurve of PSR J0034$-$0534.  Using the NE2001 electron density model of \citet{cl02}\footnote{http://rsd-www.nrl.navy.mil/7213/lazio/ne\_model/} and the measured DM gives a distance of $d=0.53\pm0.21$ kpc, assuming 40\% uncertainty due to fluctuations in the free electron density \citep{brisk}.  With this dataset it was not possible to measure the timing parallax and extract a more accurate distance measurement.  However, for a pulsar at a distance of $\sim$0.5 kpc and near the ecliptic plane the contribution of parallax to timing residuals is less than 2.4 $\mu$s \citep{hdbk} and thus well below the precision of this ephemeris.  This timing solution will be made available through the Fermi Science Support Center\footnote{http://fermi.gsfc.nasa.gov/ssc/data/access/lat/ephems/}.

\subsection{LAT Data Selection}
Analysis of LAT data was done using the \Fermi{} Science Tools\footnote{http://fermi.gsfc.nasa.gov/ssc/data/analysis/scitools/overview.html} (STs) v9r15p2. The \Fermi{} ST \emph{gtselect} was used to select events which had reconstructed sky directions within $10^{\circ}$ of the radio position of PSR J0034$-$0534, energies from 0.1 to 100 GeV, zenith angles $\leq105^{\circ}$, and spanning 2008 August 4 to 2009 September 10.  Events were required to belong to the ``Diffuse" class of events as defined under the P6\_V3 instrument response functions (IRFs), those with the highest probability of being photons \citep{LAT}.  Additionally, the \Fermi{} ST \emph{gtmktime} was used to exclude times when the rocking angle of the instrument exceeded $52^{\circ}$ and when the Earth's limb infringed upon the $10^{\circ}$ region of interest.  The events were then phase folded with the radio timing solution using the \Fermi{} plug-in\footnote{http://fermi.gsfc.nasa.gov/ssc/data/analysis/scitools/pulsar\_analysis\_appendix\_C.html\#calculatePulsePhase} now provided with the TEMPO2 software.

\section{RESULTS}

\subsection{Light Curve}
Events found within $0.8^{\circ}$ of PSR J0034$-$0534 were selected from the LAT data described in \S3.2 and tested for periodicity with the \Fermi{} ST \emph{gtptest}.  The result was an H-test \citep{htest} value of 47.7 with a chance probability of $6.8\times10^{-8}$, corresponding to a pulsed detection significance of $5.4\sigma$.  Figure 1 shows the folded light curve of these events over two rotation periods for events above 0.1 GeV and 1 GeV as well as the 300 MHz WSRT and 1.4 GHz Nan\c{c}ay radio profiles.  There is a small contribution to the pulse width in the radio profiles due to scattering; however, this contribution is much less than the bin width used in Figure 1.  The \g-ray light curve of PSR J0034$-$0534 shows two peaks which are very nearly aligned with the radio peaks.  This near-alignment of the radio and \g-ray peaks is very reminiscent of what is seen in the Crab pulsar (e.g. Abdo et al. 2010b) which is nearly aligned in radio, optical, X-ray, and \g-rays (GeV and TeV).  The $\geq$0.1 GeV light curve in Figure 1 was fit with two Lorentzians plus a constant offset, fixed at the value of the background estimate shown, which gave a reduced $\chi^{2}$ value of $\sim1.4$ indicating good agreement with the data.  Table 2 lists the peak positions ($\phi_{i}$), full width half maximum ($\rm{FWHM}_{i}$), radio to \g-ray phase lags ($\delta_{i}$), and peak separation ($\Delta$) values.  The phase lag values in Table 2 were calculated by assuming the first radio peak to be at phase 0 and estimating the second at phase 0.258, using the 324 MHz radio profile.  The phase lag in the second peak is statistically consistent with 0 while the first peak has a significant, but small, offset from the radio.  Fitting the \g-ray light curve with asymmetric Lorentzians did not improve the fit, though with more data the second peak may show significant asymmetry.  Using only events $\geq$1.4 GeV gives a pulsed detection of 3$\sigma$ while using events $\geq$2 GeV gives a pulsed detection of only 1.6$\sigma$.  This indicates that there is significant evidence for emission up to almost 2 GeV from PSR J0034$-$0534.

\subsection{Spectrum}
An unbinned maximum likelihood method (Cash 1979 and Mattox et al. 1996), using the pyLikelihood python module included with the \Fermi{} STs, was used to fit the region around PSR J0034$-$0534.  All point sources found above the background with test statistic $\geq$25 in a preliminary version of the 1FGL catalog \citep{1fgl} and within $15^{\circ}$ of PSR J0034$-$0534 were modeled with power law spectra.  The parameters of those point sources $>10^{\circ}$ from PSR J0034$-$0534 were held fixed in the fit.  The Galactic diffuse emission was modeled using the gll\_iem\_v02 map cube. The extragalactic diffuse and residual instrument background components were modeled jointly using the isotropic\_iem\_v02 template.  Both diffuse models are available for download with the \Fermi{} STs package.  The \g-ray spectrum of PSR J0034$-$0534 was modeled as both a power law and a simple exponentially cutoff power law, Equation 1 with $b\ \equiv\ 1$, in separate fits:

\begin{equation}
\frac{dN}{dE} = N_{0} \Big(\frac{E}{1\ \rm{GeV}}\Big)^{-\Gamma} \exp\Big[ -\Big(\frac{E}{E_{C}}\Big)^{b} \Big]\ .
\end{equation}

Using the likelihood ratio test, the simple exponentially cutoff power law model is preferred over a power law at the $4.5\sigma$ level.  If the emission were to come from very near the stellar surface the \g-ray spectrum could be hyper-exponentially cutoff, with $b\ >$ 1, due to pair attenuation by the magnetic field \citep{PC}.  Low-altitude emission may be expected given the near-alignment of the \g-ray and radio peaks.  Assuming a dipolar magnetic field, the predicted pair attenuation cutoff energy is larger than the curvature radiation cutoff energy for nearly all MSPs \citep{h05}; however, if the field is not dipolar the surface magnetic field could be larger and the pair attenuation cutoff energy could be low enough to dominate.  The spectrum of PSR J0034$-$0534 was also fit allowing the $b$ parameter to be free.  This fit returned a value of $b$ not statistically different from 1 and the $b\ \equiv\ 1$ model is still preferred by the likelihood ratio test, which is in agreement with the implications of the light curve modeling in \S5.1.

The \g-ray energy spectrum of PSR J0034$-$0534, with $b\ = 1$, is shown in Figure 2.  The plotted points in Figure 2 were derived from likelihood fits to each individual energy band in which it was assumed the pulsar had a power law spectrum.  The energy bands were constructed to be of equal size in log space and the last band was chosen to be that which contained the highest energy event, 6.9 GeV, found consistent with the pulsar position within the 95\% containment radius.  The best-fit parameters are given in Table 2, where the first errors are statistical and the second are systematic.  The systematic uncertainties were estimated by applying the same fitting procedures described above and comparing results using bracketing IRFs where the effective area has been perturbed by $\pm$10\% at 0.1 GeV, $\pm$5\% near 0.5 GeV, and $\pm$20\% at 10 GeV with linear extrapolations, in log space, between.  As a cross check, the \g-ray spectrum of PSR J0034$-$0534 was also fit with a binned likelihood estimator, \emph{ptlike}, which computes the photon counts in a point source weighted aperture in excess of background counts.  The \emph{ptlike} results agree with the values quoted in Table 2 within statistical and systematic errors.  Comparison with Table 1 of \citet{abdoa} shows that the spectrum of PSR J0034$-$0534 is very typical of the \g-ray MSPs known to date.  Table 2 also lists the integrated photon flux ($F$) and energy flux ($h$) from 0.1 to 100 GeV.

To search for evidence of modulation at the orbital period events were selected from the data described in \S3.2 within $5^{\circ}$ of the pulsar and with reconstructed energies between 0.1 and 10 GeV.  The modulation was assumed to be sinusoidal and a maximum likelihood method was used to fit the fraction of the average flux modulated at the orbital period.  It might be expected that an unpulsed component of emission from, for example, particle acceleration in the wind termination shock would show the strongest orbital modulation and thus choosing an off-pulse phase window would be best for this search.  However, the phase-averaged flux was chosen due to the fact that the background estimate shown in Figure 1 does not provide any strong evidence for an unpulsed component.  The likelihood was constructed by holding the spectral parameters for all sources at the phase-averaged fit values and maximizing with respect to the modulated flux and the unknown orbital phase of peak emission.  There is no evidence for modulation at the orbital period with a 95\% confidence level upper limit on any modulation of 35\% of the average flux.

The \Fermi{} ST \emph{gtobssim} was used to simulate the region around PSR J0034$-$0534 using the fit results from the maximum likelihood analysis described in \S4.2.  The simulation included all point sources within $15^{\circ}$ of the pulsar and both diffuse backgrounds but did not include the pulsar or a model of the \g-ray albedo from the Earth.  The simulation start and stop times were matched to those in the event file as closely as possible.  The same cuts described in \S4.1 were applied to the simulated data in order to get the background estimates shown in Figure 1.

\section{DISCUSSION}
\subsection{Light Curve Modeling}
In nearly all known \g-ray pulsars with radio counterparts the \g-ray peak(s) lags the radio peak(s) by at least 0.05 in phase.  The radio profile has traditionally been modeled assuming core and conal beams centered on the magnetic axis \citep{rankin} and emitted at low altitude relative to the light cylinder radius ($\rm{R}_{LC}$).  Thus, the phase lags between radio and \g-rays have been interpreted as indicating different emission altitudes for these two wavebands, with the \g-ray emission coming from outer gaps (OGs) \citep{ry} or slot gaps \citep{mh}, which have a two-pole caustic (TPC) geometry \citep{dr}, reaching much higher altitudes.  The emission peaks in OG and TPC geometries are due to the formation of caustics at phases where relativistic aberration and time-of-flight delays nearly cancel delays from magnetic field curvature on trailing field lines \citep{morini83}.

\citet{venter} modeled the light curves of the first eight \g-ray MSPs seen by \Fermi{} and observed two distinct subclasses. The radio emission for all eight was fit using a single-emission-height conal model. The \g-ray light curves of six MSPs were well modeled using an outer-magnetospheric geometry (TPC / OG model), while those of the remaining two were modeled using a pair-starved polar cap (PSPC) model in which the \g-ray emission originates from the full open-field-line volume in the pulsar magnetosphere, even up to high altitudes.  In the TPC / OG case the \g-ray profile lags the radio, while the radio lags the \g-ray pulse in the PSPC case. In contrast, PSR J0034$-$0534 is the first MSP for which the radio and \g-ray profiles are observed to be nearly aligned, providing strong evidence for co-located emission regions.  The term ``co-located'' is taken to mean that the \g-ray and radio photons are thought to be generated at similar, although not identical, locations in the magnetosphere. Fully overlapping emitting regions of the same dimensions will lead to identical light curves in these different wavebands, which is not observed. On the contrary, the different light curve shapes at different energies seem to imply that the radio emission region is smaller, and a subset of the \g-ray emission region in the case of extended, higher-altitude emission. The near-alignment of the \g-ray and radio profiles suggests two plausible configurations: either both components originate near the polar cap (PC) or both are emitted in the outer magnetosphere.  For both possibilities the altitude, extent, and degree to which the radio and \g-ray emission regions overlap are limited by the shapes of the radio and \g-ray light curves.

In the first case, relativistic effects tend to smear out the leading peaks while piling up photons in the trailing peaks, assuming constant-emissivity annular gaps near the stellar surface extending from $\sim0.1$ to 0.2R$_{LC}$.  This is contrary to what is observed in the \g-ray light curve.  It is also difficult to reproduce the symmetric radio peaks assuming emission near the PC, even for emission at the stellar surface.  If the emission regions are too high in altitude (or their extent is too large) the relativistic effects would be boosted (since the corresponding phase shifts due to these effects scale as $\sim -r/$R$_{LC}$) and the peaks would be even more asymmetric, with a sharp trailing peak. Conversely, too little overlap would negate the phase alignment.

There is however a special solution that is an exception to the rule for the low-altitude models. Usually, the peak separation ($\Delta$) is defined as the difference in phase between the leading and trailing peak positions. In certain circumstances, the leading (smeared-out) and trailing (sharp) peaks appear ``inverted'' to an observer at a particular observer angle $\zeta$, the angle between the rotation axis and the observer's line-of-sight. The trailing peak would be interpreted as the first peak, and the leading peak as the second one, since the peak separation between the first and second peaks ($1-\Delta$ in normalized phase) would be smaller than 0.5. If the inclination angle between the rotation and magnetic axes ($\alpha$) is small enough ($\lesssim10^{\circ}$), one may find solutions with relative peak intensities and separations that fit the data quite well. More details are provided in \citet{vh2}. Although this is a valid solution, it is a less likely geometry due to the very specific choices of $\alpha$ and $\zeta$ that are needed to obtain a good fit. Unfortunately, the only radio polarization measurement of PSR J0034$-$0534 to date was inconclusive and unable to put any constraints on the pulsar geometry \citep{pol}.

A co-located outer-magnetospheric origin of both the radio and \g-ray radiation seems more likely.  If this is the case, PSR J0034$-$0534 is the first example of yet another MSP subclass, distinct from the two \g-ray MSP subclasses found by \citet{venter}.  \citet{manch} has suggested that radio emission for young and millisecond pulsars may be generated in the outer magnetosphere, close to the \g-ray emission region, an idea which \citet{wj} explored by comparing the characteristics of pulsars with high and low $\dot{\rm{E}}_{SD}$ values in the Parkes multi-beam surveys.  Additionally, \citet{drd} have proposed a model in which the radio emission for MSPs with symmetric, double-peaked radio features, like what is observed in PSR J0034$-$0534, arise from curvature radiation from plasma streams with a non-negligible range of emission altitudes.

Figure 3 shows model \g-ray and radio light curves generated in the context of geometric `limited TPC and OG models', i.e., TPC / OG models with limited extent (along the B-field) of the emission regions. The minimum and maximum emission radii for both the radio and \g-ray components are constrained by the light curve shapes for each band: excluding lower-altitude emission lowers the off-pulse `shoulder emission' (i.e., emission between consecutive profiles at phases $\sim 0.4-0.9$), and `bridge emission' (inter-peak emission at phases $\sim 0.0-0.3$) while giving rise to sharper, more separated peaks.  On the other hand, including more of the higher-altitude emission boosts the leading peak and eventually widens (and to a lesser extent boosts) the trailing peak.  A more detailed treatment of these `limited TPC / OG models' can be found in \citet{vh2}.

Given these constraints, the radio and \g-ray light curves are well modeled using $\alpha\ =\ 30^{\circ}$, $\zeta\ =\ 70^{\circ}$, and a transverse gap width of $w\ =\ 0.05$ (the fractional angular width starting at the PC rim, and normalized to the colatitude of the rim).  For the limited TPC models in Figure 3 the \g-ray emission region extends (in radius) from 0.12R$_{LC}$ to 0.9R$_{LC}$ (i.e., starting at the stellar surface) while the radio emission region extends from 0.6R$_{LC}$ to 0.8R$_{LC}$.  The ranges of emission radii for the limited OG models in Figure 3 are the same as those for the TPC model with the added caveat that the emission cannot extend below the null charge surface.  The \g-ray model light curves provide a reasonable fit to the data, although they all somewhat over predict the bridge emission as compared to the prediction of the radio curves. The agreement of the observed and modeled profiles strongly suggests that both the \g-ray and radio pulses originate in the outer magnetosphere with the peaks formed by caustics.  It is of interest to note that there is a slight excess of events in the \g-ray light curve immediately preceding the first peak which is currently insignificant.  This feature is reproduced with the TPC model but not with the OG model, since it is produced by emission below the null charge surface.  With more data, the presence or absence of this feature should become statistically better defined and will be useful in further discriminating between emission models.

\subsection{Efficiency}
The total luminosity in \g-rays from PSR J0034$-$0534 can be calculated using Equation 2:
\begin{equation}
L_{\gamma}\ =\ 4\pi f_{\Omega} h d^{2}\ .
\end{equation}
The correction factor $f_{\Omega}$ depends on the viewing geometry and beaming angle of the pulsar and is typically $\sim$1 for outer-magnetospheric emission models.  Using an approach similar to \citet{venter} and \citet{watters} yields $f_{\Omega} = 0.74$ for TPC models and $f_{\Omega} = 0.45$ for OG models, assuming the same geometry used in \S5.1.  Using the derived energy flux from Table 2 and the DM-derived distance from \S3.1 yields $L_{\gamma} = 4.7(2.9)\pm3.8(2.3)\times10^{32}\ \rm{erg}\ \rm{s}^{-1}$ for the TPC(OG) model.

Once the luminosity is known, the efficiency with which PSR J0034$-$0534 converts spin-down energy into \g-rays can be calculated using Equation 3:
\begin{equation}
\eta_{\gamma}\ =\ \frac{L_{\gamma}}{\dot{\rm{E}}_{SD}}
\end{equation}
where $\dot{\rm{E}}_{SD}$ is the rate at which energy is lost due to the pulsar spinning down.  This gives an efficiency of $\eta_{\gamma} = 0.03(0.02)\pm0.03(0.02)$ for the TPC(OG) model.  The derived efficiency for PSR J0034$-$0534 is on the low end of the efficiencies quoted in Table 1 of \citet{abdoa}, which were calculated assuming $f_{\Omega} = 1$, and lower than the upper limit of $\sim0.1$ estimated for MSPs in the GC 47 Tucanae \citep{abdob}.  The large uncertainties of $L_{\gamma}$ and $\eta_{\gamma}$ are primarily due to the uncertainty of the DM-derived distance in the proper motion.  A parallax measurement would greatly increase the precision of these values.

\section{CONCLUSIONS}
The MSP PSR J0034$-$0534 is the ninth \g-ray MSP detected with the \Fermi{} LAT.  The \g-ray spectral properties of PSR J0034$-$0534 are similar to those of other \g-ray MSPs detected by \Fermi{} thus far; however, PSR J0034$-$0534 is (so far) unique in that it is the only known \g-ray MSP in which the radio and \g-ray peaks are very nearly aligned in phase.  The near phase alignment can be interpreted as providing strong evidence for co-located emission regions.  Within the context of geometric TPC and OG models, the radio and \g-ray light curves are well modeled by requiring that both emission regions be significantly extended in altitude.  This implies that both the radio and \g-ray emission peaks are a result of caustic formation.  A similar radio geometry was required to model the Crab \g-ray and radio light curves \citep{hcrab}, which are also coincident in phase.  As the \Fermi{} mission continues and more \g-ray MSPs are detected this new and interesting MSP subclass may be further populated.

\acknowledgements
\emph{Acknowledgements}\\
The \Fermi{} LAT Collaboration acknowledges generous ongoing support from a number of agencies and institutes that have supported both the development and the operation of the LAT as well as scientific data analysis.  These include the National Aeronautics and Space Administration and the Department of Energy in the United States; the Commissariat \`a l'Energie Atomique and the Centre National de la Recherche Scientifique / Institut National de Physique Nucl\'eaire et de Physique des Particules in France; the Agenzia Spaziale Italiana and the Istituto Nazionale di Fisica Nucleare in Italy; the Ministry of Education, Culture, Sports, Science and Techology (MEXT), High Energy Accelerator Research Organization (KEK) and Japan Aerospace Exploration Agency (JAXA) in Japan; and the K. A. Wallenberg Foundation, the Swedish Research Council and the Swedish National Space Board in Sweden.

Additional support for science analysis during the operations phase is gratefully
acknowledged from the Istituto Nazionale di Astrofisica in Italy and the Centre National d'\'Etudes Spatiales in France.

The Nan\c{c}ay Radio Observatory is operated by the Paris Observatory, associated with the French Centre National de la Recherche Scientifique (CNRS).

The Westerbork Synthesis Radio Telescope is operated by Netherlands Foundation for Radio Astronomy, ASTRON.

\begin{deluxetable}{l l}
\tablecaption{PSR J0034$-$0534 Timing Parameters}
\startdata
\underline{Measured Parameters}\tablenotemark{a} \\
P (ms) & 1.8771818845850(2)\\
$\dot{\rm{P}}\ (10^{-21})$ & 4.966(1)\\
$\mu$ (mas $\rm{yr}^{-1}$)\tablenotemark{b} & 31(9)\\
\\
\underline{Derived Parameters}\tablenotemark{c} \\
$d$ (kpc)\tablenotemark{d} & 0.53 $\pm$ 0.21\\
$\dot{\rm{P}}_{\rm{corr}} (10^{-21})$ & 2.63 $\pm$ 1.64\\
$\rm{B}_{surf}$ ($10^{7}$ G) & $7.12\ \pm\ 2.22$\\
$\dot{\rm{E}}_{SD} (10^{34}\ \rm{erg}\ \rm{s}^{-1})$ & $1.57\ \pm\ 0.98$\\
$\rm{B}_{LC}$ ($10^{4}$ G) & $9.89\ \pm\ 3.09$\\
\enddata
\tablenotetext{a}{Values in parentheses are rms errors on the last digit.}
\tablenotetext{b}{\citet{hobbs05}.}
\tablenotetext{c}{Derived using $\dot{\rm{P}}_{\rm{corr}}$, which accounts for Shklovskii effect.}
\tablenotetext{d}{DM-derived using the NE2001 model of \citet{cl02}.}
\end{deluxetable}

\begin{deluxetable}{l r}
\tablecaption{PSR J0034$-$0534 \g-ray Parameters}
\startdata
\underline{Light Curve Parameters}\tablenotemark{a}\\
$\phi_{1}$ & $-0.027\ \pm\ 0.008$\\
FWHM$_{1}$ & $0.066\ \pm\ 0.019$\\
$\delta_{1}$ & $-0.027\ \pm\ 0.008$\\
$\phi_{2}$ & $0.247\ \pm\ 0.013$\\
FWHM$_{2}$ & $0.106\ \pm\ 0.038$\\
$\delta_{2}$ & $0.011\ \pm\ 0.013$\\
$\Delta$ & $0.274\ \pm\ 0.015$\\
\\
\underline{Spectral Parameters}\tablenotemark{b}\\
$\rm{N}_{0}\ (10^{-9}\ \rm{cm}^{-2}\ \rm{s}^{-1}\ \rm{GeV}^{-1})$ & $6.9\ \pm\ 1.8\ \pm\ 0.2$\\
$\Gamma$ & $1.5\ \pm\ 0.2\ \pm\ 0.1$\\
$\rm{E}_{C}$ (GeV) & $1.7\ \pm\ 0.6\ \pm\ 0.1$\\
$F\ (10^{-8}\ \rm{cm}^{-2}\ \rm{s}^{-1})$ & $2.7\ \pm\ 0.5\ \pm\ 0.4$\\
$h\ (10^{-11}\ \rm{erg}\ \rm{cm}^{-2}\ \rm{s}^{-1})$ & $1.9\ \pm\ 0.2\ \pm\ 0.1$\\
\enddata
\tablenotetext{a}{Errors are statistical.  Parameters are described in \S4.1.}
\tablenotetext{b}{First errors are statistical, second are systematic.  Parameters are described in \S4.2}
\end{deluxetable}

\begin{figure}
\epsscale{0.9}
\plotone{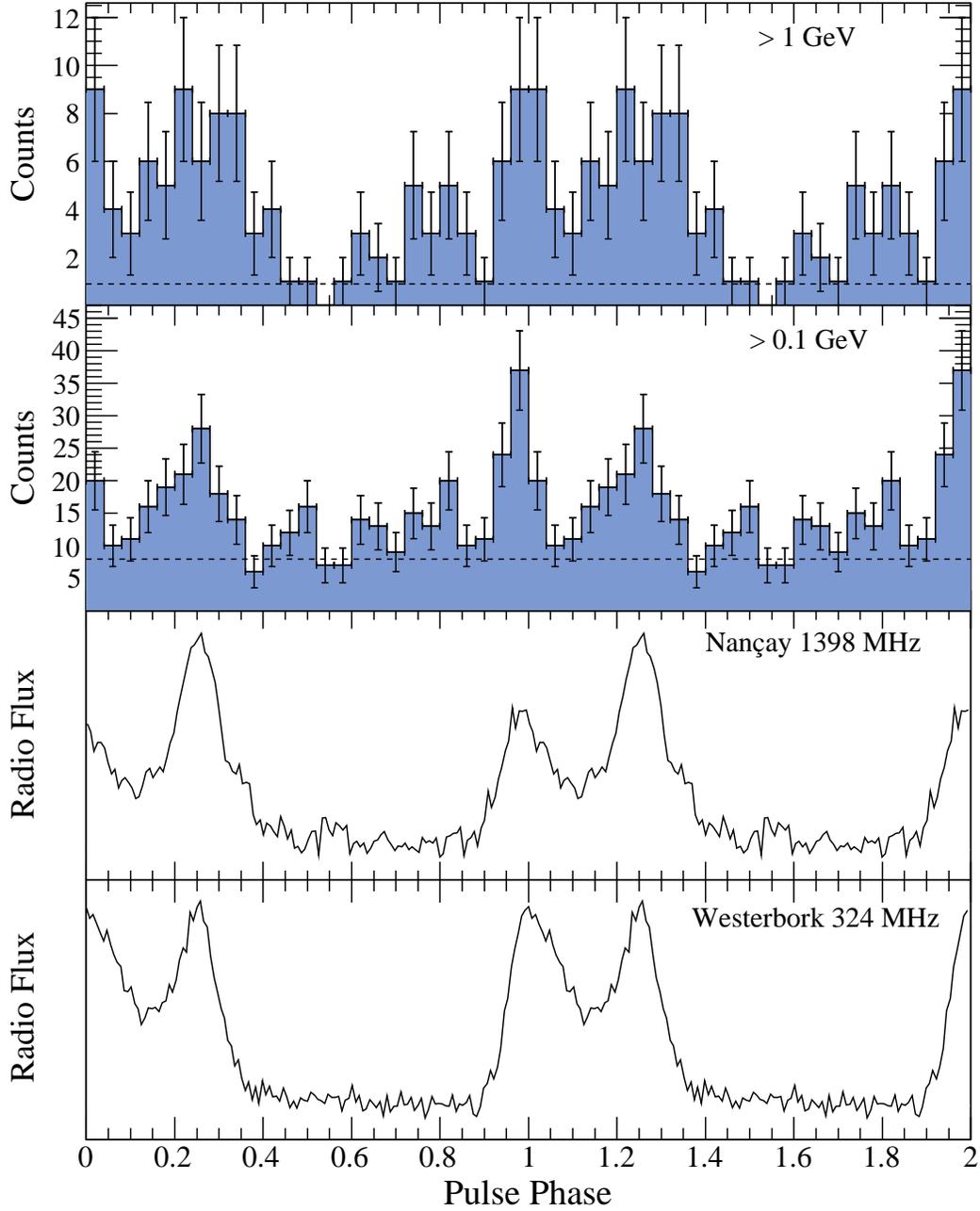}
\caption{The top two panels show the phase-folded light curve of PSR J0034$-$0534 for LAT events above 1 GeV and above 0.1 GeV within $0.8^{\circ}$ of the radio position. \g-ray light curves are shown across two rotations with 25 bins per rotation.  The dashed horizontal lines correspond to the background levels estimated from the simulation described in \S4.2. The bottom two panels show the Nan\c{c}ay and WSRT radio profiles; the vertical axes are in arbitrary units.}
\end{figure}

\begin{figure}
\plotone{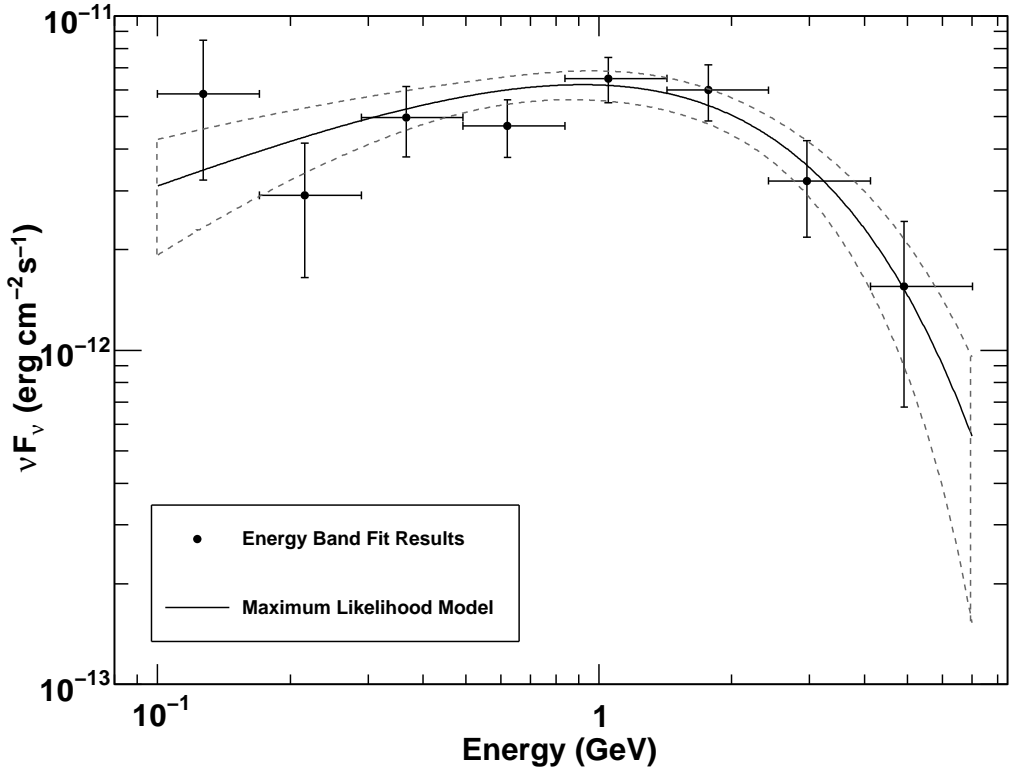}
\caption{Phase-averaged \g-ray energy spectrum of PSR J0034$-$0534.  Plotted points are from likelihood fits to individual energy bands where the pulsar is modeled as a power law, solid black line is the maximum likelihood model from fitting the full energy range, dashed gray lines are the $1\sigma$ errors on the model.  All sources described in \S4.2 were modeled but only the parameters of those within $6^{\circ}$ were left free in the fits.  For each energy band the pulsar was found above the background with a test statistic of at least 6, $\geq2\sigma$ for two degrees of freedom.}
\end{figure}

\begin{figure}
\plotone{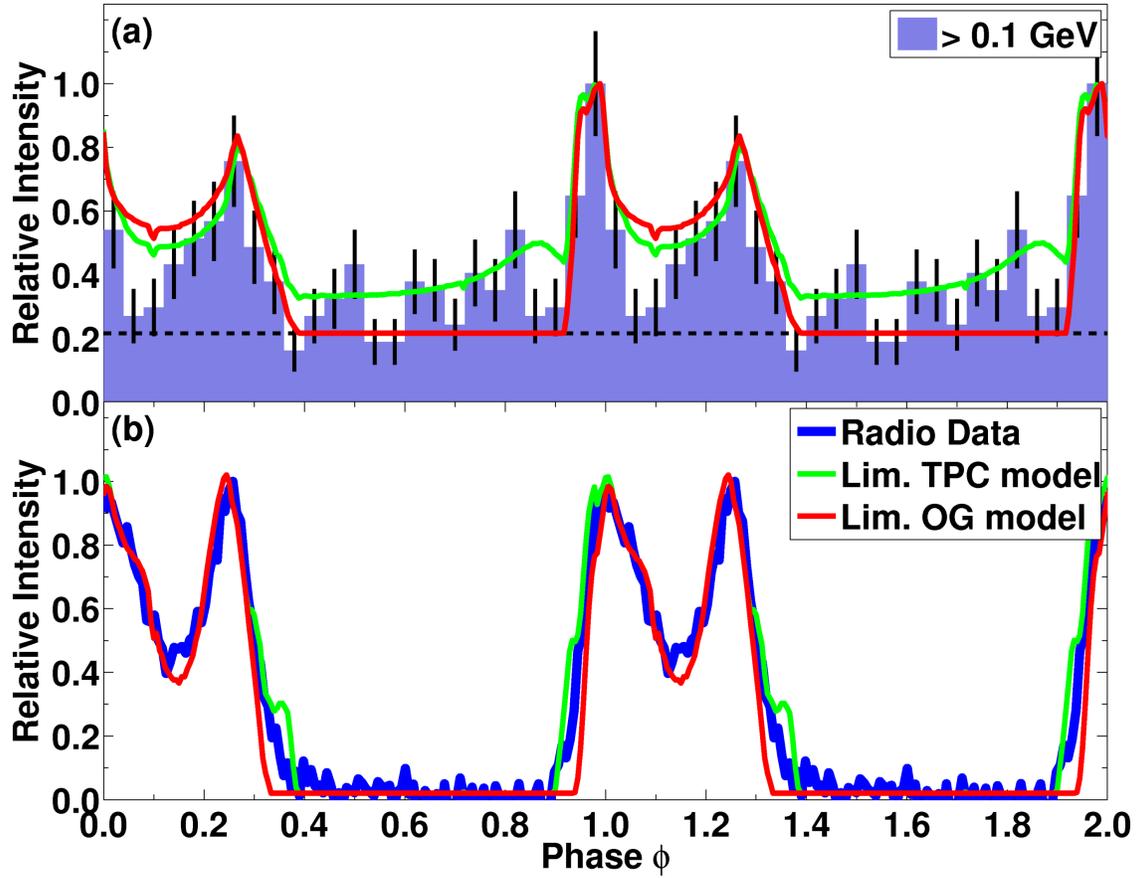}
\caption{\emph{Top:} \g-ray data and modeled light curves.  \emph{Bottom:} WSRT 324 MHz radio profile and modeled light curves.  All modeled light curves were made using $\alpha\ =\ 30^{\circ}$, $\zeta\ =\ 70^{\circ}$, and $w\ =\ 0.05.$  The extent of the limited TPC and OG models are given in \S5.1.}
\end{figure}

\maketitle

\end{document}